\documentclass[11pt]{article}
\usepackage[preprint]{acl}
\usepackage{times,latexsym}
\usepackage[T1]{fontenc}
\usepackage[utf8]{inputenc}
\usepackage{microtype,inconsolata}
\usepackage{graphicx,booktabs,amsmath,amssymb,tabularx,array,xcolor}
\usepackage{tikz,pgfplots}
\pgfplotsset{compat=1.18}\usetikzlibrary{patterns}
\usetikzlibrary{arrows.meta,positioning,fit}
\usepackage{algorithm,algpseudocode,placeins,float}
\usepackage{listings}
\lstdefinestyle{prompt}{basicstyle=\ttfamily\footnotesize,breaklines=true,columns=fullflexible,keepspaces=true,literate={×}{{$\times$}}1,frame=single,rulecolor=\color{black!35},showstringspaces=false,xleftmargin=3pt,xrightmargin=3pt}
\newcommand{\pending}{\textemdash}

\title{BioPRM: Step-Level Consistency Verification of Biological Protocols}
\author{
  Bowen Shao$^{1,2}$ \quad Ziyang Ling$^{1,2}$ \quad
  Ronald Xu$^{1,2,*}$ \quad Mingzhai Sun$^{1,2,*}$ \\
  {\normalfont\small $^{1}$Suzhou Institute for Advanced Research,} \\
  {\normalfont\small University of Science and Technology of China} \\
  {\normalfont\small $^{2}$School of Biomedical Engineering, Division of Life Sciences and Medicine,} \\
  {\normalfont\small University of Science and Technology of China, Hefei, Anhui, China} \\
  {\normalfont\small $^{*}$Corresponding authors: \texttt{mingzhai@ustc.edu.cn}, \texttt{xu@ustc.edu.cn}}
}
\begin{document}
\maketitle
\raggedbottom
\begin{abstract}
Process reward models (PRMs) provide fine-grained supervision for assessing intermediate steps in language-model outputs, but their application to biological protocols presents distinct challenges. Protocol validity depends on both the appropriateness of individual operations and their consistency with the experimental objective and surrounding procedure, while experimental outcomes are typically unavailable during verification. We introduce BioPRM, a contextual process reward model for identifying procedural inconsistencies in structured biological protocols. We construct a supervision corpus from SciRecipe and protocols collected from protocols.io and Bio-protocol using LLM-assisted construction and source-specific review by personnel with biological expertise, with labels for step correctness, protocol correctness, and conditional error types. Built on a protocol-adapted Llama-3.1-8B backbone, BioPRM combines step-specific representations with a full-protocol readout and uses two-stage training: multi-task supervised verification followed by continuation with matched-pair ranking. On a test set of 2,649 protocols and 11,701 steps, BioPRM achieves 84.17\% step average precision and 76.92\% step error F1, together with 93.51\% protocol average precision and 84.12\% protocol error F1. These results demonstrate the feasibility of adapting process-supervised verification to biological protocols, providing step-level error localization and diagnosis alongside protocol-level assessment.
\end{abstract}

\section{Introduction}
Biological experimental protocols translate research objectives into ordered instructions for manipulating materials and collecting observations. Language models increasingly assist this process: BioPlanner studies protocol reconstruction through pseudocode \citep{odonoghue-etal-2023-bioplanner}, while Thoth supports structured protocol generation and is introduced alongside the SciRecipe dataset \citep{sun2026scirecipe}. However, a plausible procedure can omit a prerequisite operation, use an unsuitable material, or rely on a condition that an earlier step has not established. PRISM documents prerequisite failures in generated laboratory workflows, and BioProBench evaluates protocol ordering and error assessment \citep{hsu2026prism,liu2025bioprobench}. Such findings motivate checking candidate procedures before execution.

Verification presents a language-understanding problem as well as a domain-knowledge problem. The meaning and validity of an instruction can depend on entities and conditions introduced elsewhere in the protocol. An operation may be appropriate in isolation but incompatible with the preceding procedure. Meanwhile, experimental outcomes are often unavailable at verification time, and disagreement with one reference protocol cannot establish invalidity when legitimate alternatives exist. We study text-based verification conditioned on the experimental request and the complete candidate procedure.

Prior work already addresses biological protocol errors and dependencies. BioLP-bench and BioProBench evaluate error identification, while hierarchical protocol representations and BioProAgent incorporate explicit dependencies or constraints \citep{ivanov2024biolp,liu2025bioprobench,shi2025hierarchical,liu2026bioproagent}. We study a learned verifier that integrates complete-protocol context with fine-grained supervision, enabling localized judgments and typed feedback. Our experiments examine how global representations, auxiliary diagnostic labels, and paired supervision contribute to this biological-protocol verification task.

Process reward models (PRMs) learn from intermediate judgments rather than only a sequence-level outcome. They have been studied in mathematical reasoning, clinical-note verification, and scientific tool use \citep{lightman2023verify,wang-etal-2025-process,zhao2026sciprm}. Protocol steps provide a different supervision unit: an experimental operation interpreted against a query and other operations. ProcessBench motivates direct evaluation of step-verification behavior and comparisons with prompted critics \citep{zheng2025processbench}.

We introduce \textbf{BioPRM}, a contextual verifier for structured biological protocols. We refer to its training data as the \emph{BioPRM supervision corpus}. Reference protocols are drawn from SciRecipe and procedures collected from protocols.io and Bio-protocol. LLM-assisted construction produces protocol, step, and conditional error-type labels, with independent manual review of the self-collected data. BioPRM combines representations of individual step boundaries with a readout of the completed protocol. Stage I jointly trains three prediction heads; Stage II retains the classification objectives while adding target-step ranking between strictly matched reference and perturbed protocols. The main reported model is the validation-selected two-stage checkpoint, and Stage I is retained as an explicit comparator.

Our contributions are twofold. First, we formulate and implement joint protocol assessment, designated-error localization, and type diagnosis using fine-grained process supervision. Second, we analyze the verifier and its training components on constructed errors. Global readout improves the reported Stage I detection results, while type supervision improves step detection and supplies diagnostic predictions in the reported comparison. The matched-pair variant achieves higher scores than the reported classification-only continuation, and comparison with Stage I characterizes the effect of continued training across detection and diagnosis metrics. Corpus construction and reproducibility details are provided in the appendices.

\section{Related Work}
\paragraph{Protocol representation and verification.}
Process Execution Graphs and material state transfer graphs model cross-sentence arguments and temporal or causal action dependencies \citep{tamari-etal-2021-process,kulkarni-etal-2021-learning}. BioPlanner and SciRecipe address protocol planning and structured generation \citep{odonoghue-etal-2023-bioplanner,sun2026scirecipe}. BioLP-bench and BioProBench already evaluate protocol error identification, and BenchBench-Protocol studies modifications derived from scientists' actual protocol adaptations \citep{ivanov2024biolp,liu2025bioprobench,sivakumar2026benchbench}. We focus on a verifier that predicts a judgment at every annotated step, an auxiliary error type, and a separate protocol judgment. Controlled perturbations provide aligned correctness and diagnostic targets for studying these predictions.

\paragraph{Process supervision and diagnosis.}
Process supervision has been studied in mathematical reasoning, including automatic step labeling and planning-based supervision \citep{lightman2023verify,wang2024mathshepherd,pisano-navigli-2026-process}. Clinical PRM is a close methodological precedent: it defines domain-specific steps and errors, uses LLM-generated corruptions with manual inspection, and trains a verifier for clinical notes \citep{wang-etal-2025-process}. BioPRM assesses experimental operations and procedural dependencies in structured biological protocols. Our corpus review is tailored to structured experimental procedures. PathFinder-PRM studies error-typed supervision, while SCI-PRM and Med-PRM extend process verification to scientific and medical tasks \citep{pala2025pathfinder,zhao2026sciprm,yun-etal-2025-med}. BioPRM contributes a biological-protocol supervision corpus, a contextual multi-granularity verifier, and an empirical analysis of its training components.

\paragraph{Alternative verification signals.}
Outcome-supervised models can provide sequence judgments, and implicit process rewards can be derived from outcome labels under a particular reward parameterization \citep{yuan2024free}. ProcessBench and PRMBench expose limitations of learned step verifiers and motivate comparison with prompted critics \citep{zheng2025processbench,song-etal-2025-prmbench}. External verification offers another route: CLAIRify checks structured programs; LLM-Modulo combines language-model planning with model-based verifiers; hierarchical protocol representations, BioProAgent, and PRISM use domain constraints or simulated execution \citep{skreta2023clairify,kambhampati2024llmmodulo,shi2025hierarchical,liu2026bioproagent,hsu2026prism}. BioPRM provides text-based consistency assessment and localized diagnostic feedback for reviewing candidate procedures.

\section{Problem Definition}
Given an experimental query $q$ and a completed candidate protocol $P=(s_1,\ldots,s_n)$, we identify designated procedural violations. Each step is serialized from structured action, object, and parameter fields. The query specifies the objective and stated constraints. A protocol may contain no annotated violation or several; valid alternatives to a reference should remain acceptable.

A process verifier returns
\begin{equation}
V_\theta(q,P)=
\left(p^o,\{p_i,\boldsymbol{\pi}_i\}_{i=1}^{n}\right),
\label{eq:verifier}
\end{equation}
where $p_i$ and $p^o$ are step and protocol error scores, respectively. The conditional distribution $\boldsymbol{\pi}_i$ covers ACTION, OBJECT, PARAMETER, and STATE errors. These categories describe inappropriate operations, incompatible entities, invalid conditions, and unmet procedural dependencies (Appendix~\ref{app:dataset}). Stored correctness labels $c_i,c^o$ use 1 for CORRECT and 0 for ERROR; error-positive targets are $y_i=1-c_i$ and $y^o=1-c^o$.

Labels define designated violation locations and their primary diagnostic categories. Step-level evaluation measures agreement with these targets, supporting a consistent definition of localization across protocols. At inference the verifier receives only $q$ and $P$; reference protocols, labels, and perturbation metadata are excluded.

\section{Method}
\label{sec:method}
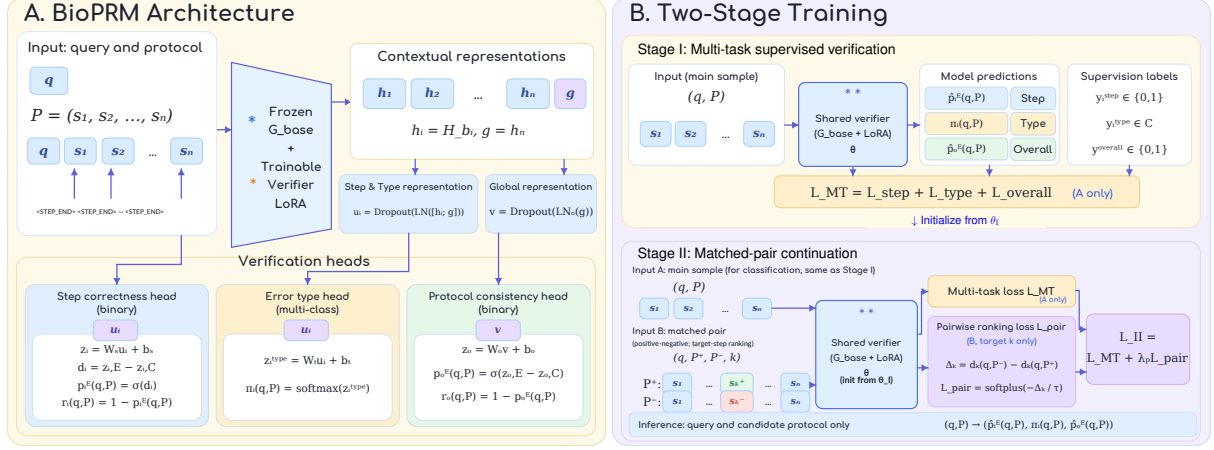
\begin{figure*}[t]
\centering
\resizebox{\textwidth}{!}{%
\begin{tikzpicture}[x=1pt,y=1pt]
\node[anchor=south west,inner sep=0] at (0,0) {\includegraphics[width=1536pt]{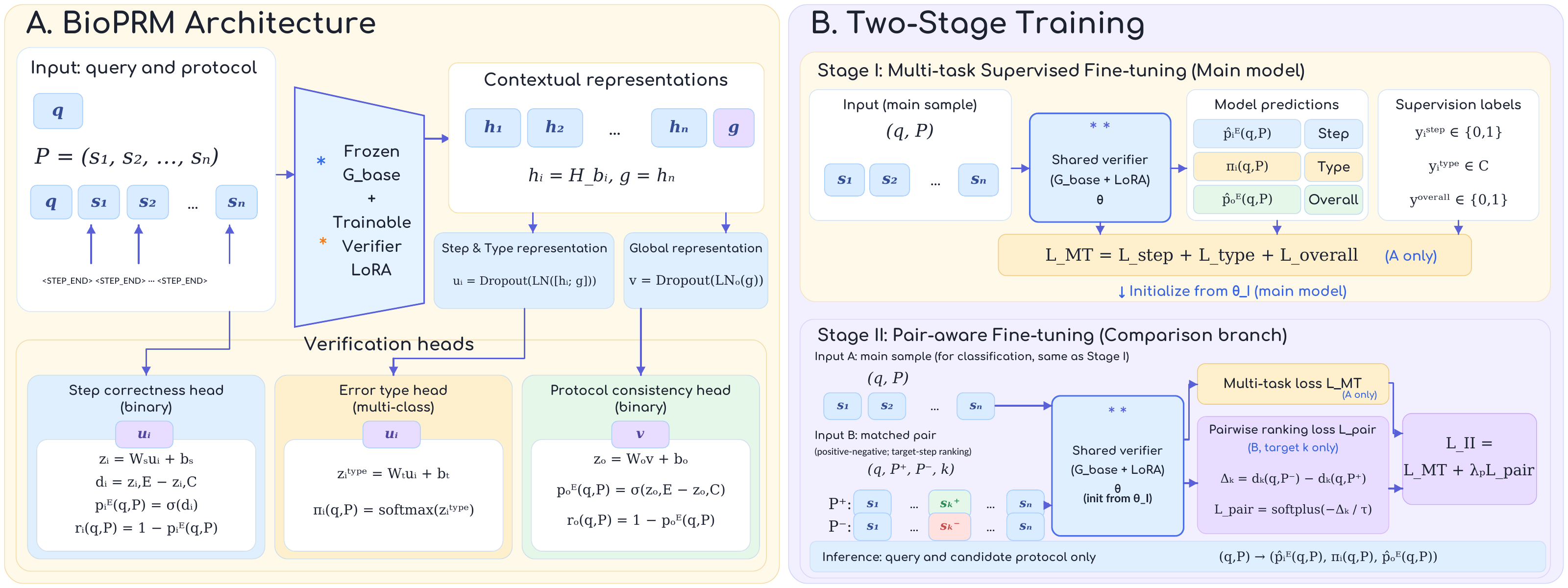}};
\definecolor{stageoneband}{HTML}{FFF9E8}
\definecolor{stagetwoband}{HTML}{F2EFFF}
\fill[stageoneband] (799,495) rectangle (1330,519);
\node[anchor=west,font=\sffamily\fontsize{18}{21}\selectfont] at (801,506) {Stage I: Multi-task supervised verification};
\fill[stagetwoband] (799,235) rectangle (1340,258);
\node[anchor=west,font=\sffamily\fontsize{18}{21}\selectfont] at (801,247) {Stage II: Matched-pair continuation};
\fill[stageoneband] (1093,281) rectangle (1322,300);
\node[anchor=center,text=blue,font=\sffamily\fontsize{14.4}{17}\selectfont] at (1208,290) {$\downarrow$ Initialize from $\theta_{\rm I}$};
\end{tikzpicture}}
\caption{Verifier architecture and two-stage training. Stage I trains the initial verifier; Stage II continues its weights with classification supervision and target-step ranking. The selected Stage-II checkpoint is the primary BioPRM model; Stage I is a separate comparator. Type supervision applies only to erroneous steps. The diagram's logits $(z_i,z_i^{\rm type},z_o)$ correspond to $(a_i,t_i,a^o)$ in the text; $p_i^E=p_i$ and $p_o^E=p^o$, with complementary correctness scores $r=1-p^E$. Diagram inputs A/B denote main records/pairs, respectively; the corresponding batch sets are $\mathcal B/\mathcal Q$ in Eq.~\ref{eq:total}. The displayed pair loss is per pair; Eq.~\ref{eq:pair} averages it. Inference requires only the query and candidate protocol.}
\label{fig:framework}
\end{figure*}

BioPRM combines a contextual encoder, three prediction heads, and two training stages (Figure~\ref{fig:framework}). Stage I learns classification targets; Stage II continues the selected weights with matched-pair ranking while retaining classification supervision.

\subsection{Contextual Protocol Encoding}
The backbone is initialized from $G_{\rm base}$, a Llama-3.1-8B-Instruct model previously adapted for structured protocol generation. Fresh verifier LoRA adapters, normalization layers, and prediction heads are trained for verification. This preceding generator adaptation is initialization history, separate from the two verifier phases.

We serialize the query and ordered instructions as $T(q,P)$. Each step ends with the marker string \texttt{<STEP\_END>}; tokenizer offsets locate its final token $b_i$. It is not added as a single vocabulary token. From the final hidden states
$H=F_\theta(T(q,P))$, we extract
\begin{equation}
h_i=H_{b_i},\qquad g=h_n.
\end{equation}
The causal state $h_i$ contains the prefix through $i$; $g$ has access to the completed protocol. Their processed features are
\begin{align}
u_i&=\operatorname{Dropout}
       \bigl(\operatorname{LN}_s([h_i;g])\bigr),\\
v&=\operatorname{Dropout}
       \bigl(\operatorname{LN}_o(g)\bigr).
\label{eq:fusion}
\end{align}
The causal state supplies prefix information, while the shared readout exposes the completed protocol. For the final step, $h_n=g$, so the two concatenated components coincide. The representation combines prefix and completed-protocol information in a shared contextual feature space.

\begin{algorithm*}[t]
\caption{Two-Stage Training and Validation Selection for BioPRM}
\label{alg:train}
\footnotesize
\begin{algorithmic}[1]
\Require Training records $D_{\rm train}$, validation records $D_{\rm val}$; frozen backbone $G_{\rm base}$;
\Statex Stage-I budget $T_1$; Stage-II segment budgets $(T_{2,1},\ldots,T_{2,J})$; coefficient $\lambda_p$ and temperature $\tau$
\Ensure Stage-I verifier $\theta_{\rm I}$, two-stage verifier $\theta_{\rm II}$, decision thresholds $t_s,t_o$
\State Initialize trainable $\theta$: LoRA adapters, readout normalization layers, and prediction heads
\State Freeze backbone weights
\Statex \textit{Stage I: multi-task supervised verification}
\State Initialize the Stage-I optimizer and learning-rate scheduler
\For{each optimizer update up to $T_1$}
 \State Sample classification records $\mathcal B$ from $D_{\rm train}$
 \State Encode query and protocol; obtain step and protocol predictions
 \State Compute $\mathcal L_{\rm MT}(\mathcal B)=\mathcal L_s(\mathcal B)+\mathcal L_t(\mathcal B)+\mathcal L_o(\mathcal B)$
 \State Update $\theta$ using accumulated gradients of $\mathcal L_{\rm MT}(\mathcal B)$
 \State At evaluation intervals, evaluate on $D_{\rm val}$ and save checkpoints
\EndFor
\State Select $\theta_{\rm I}$ using VAL Step AP under the recorded guard/AP-fallback rule (Appendix~\ref{app:selection})
\Statex \textit{Stage II: classification with matched-pair ranking}
\State Construct eligible matched pairs $\mathcal M$ from $D_{\rm train}$ only
\State Load $\theta\gets\theta_{\rm I}$; initialize candidate set $\mathcal C_{\rm II}\gets\varnothing$
\For{each continuation segment $j=1,\ldots,J$}
 \State Initialize a fresh optimizer and learning-rate scheduler
 \For{each optimizer update up to $T_{2,j}$}
  \State Sample classification records $\mathcal B$ from $D_{\rm train}$ and matched pairs $\mathcal Q$ from $\mathcal M$
  \State Compute $\mathcal L_{\rm MT}(\mathcal B)$ on classification records only
  \State For each $(q,P^+,P^-,k)\in\mathcal Q$, compute $\Delta_k=d_k(q,P^-)-d_k(q,P^+)$
  \State Compute $\mathcal L_{\rm pair}(\mathcal Q)=|\mathcal Q|^{-1}\sum_{(q,P^+,P^-,k)\in\mathcal Q}\operatorname{softplus}(-\Delta_k/\tau)$
  \State Update $\theta$ using accumulated gradients of $\mathcal L_{\rm MT}(\mathcal B)+\lambda_p\mathcal L_{\rm pair}(\mathcal Q)$
  \State At evaluation intervals, evaluate on $D_{\rm val}$ and add saved checkpoints to $\mathcal C_{\rm II}$
 \EndFor
 \State Retain segment-end weights for the next segment
\EndFor
\State Select $\theta_{\rm II}$ from $\mathcal C_{\rm II}$ by maximum validation Step AP
\Statex \textit{Decision thresholds}
\State Using $\theta_{\rm II}$, obtain step and protocol error scores on $D_{\rm val}$
\State Select $t_s,t_o$ independently to maximize the corresponding ERROR-class F1; choose the higher threshold on ties
\State \Return $\theta_{\rm I},\theta_{\rm II},t_s,t_o$
\end{algorithmic}
\end{algorithm*}

\subsection{Multi-Granularity Verification}
Step and protocol heads each produce two logits, ordered as ERROR (index 0) and CORRECT (index 1). An auxiliary head predicts the $K=4$ diagnostic categories:
\begin{align}
a_i&=W_su_i+b_s\in\mathbb R^2,\\
t_i&=W_tu_i+b_t\in\mathbb R^K,\\
a^o&=W_ov+b_o\in\mathbb R^2.
\end{align}
The backbone hidden size is 4,096. Concatenation gives 8,192-dimensional Step and Type inputs; Overall receives a 4,096-dimensional input. With error logit differences
\begin{equation}
d_i=a_{i,\mathrm E}-a_{i,\mathrm C},\qquad
d^o=a^o_{\mathrm E}-a^o_{\mathrm C},
\end{equation}
predictions are
\begin{equation}
p_i=\sigma(d_i),\quad p^o=\sigma(d^o),\quad
\boldsymbol{\pi}_i=\operatorname{softmax}(t_i).
\end{equation}
Type supervision applies to annotated errors and supplies a primary diagnostic category. The directly supervised Overall head complements these localized predictions with a protocol-level judgment.

\subsection{Stage I: Multi-Task Supervised Verification}
Here, $\mathcal B$ denotes the active classification records in a device microbatch. Gradient accumulation and distributed training combine gradients across microbatches to obtain the effective batch size. Let $m_{bi}$ be a valid step mask. We use error-positive targets $y_{bi}=1-c_{bi}$ and average valid steps within each protocol before averaging protocols:
\begin{equation}
\mathcal L_s=\frac1{|\mathcal B|}\sum_{b\in\mathcal B}
\frac{\sum_i m_{bi}\ell_{\rm BCE}(d_{bi},y_{bi})}
{\max(1,\sum_i m_{bi})}.
\end{equation}
Here $\ell_{\rm BCE}$ is binary cross-entropy with logits. This reduction weights protocols independently of their step counts.

Let $E_b$ contain valid erroneous steps and $\mathcal B_E=\{b\in\mathcal B:|E_b|>0\}$. The diagnostic loss is
\begin{equation}
\mathcal L_t=
\frac1{|\mathcal B_E|}\sum_{b\in \mathcal B_E}
\frac{\sum_{i\in E_b}-w^t_{\kappa_{bi}}
\log\pi_{bi,\kappa_{bi}}}{|E_b|},
\label{eq:type_loss}
\end{equation}
where $\kappa_{bi}$ is the diagnostic target and $w^t$ its class weight. The loss is zero when $\mathcal B_E$ is empty. Diagnostic losses are averaged over valid erroneous steps within each protocol and then over protocols containing such targets. The denominator counts valid error steps, rather than the sum of class weights.

For main records with valid protocol targets, $\mathcal B_o$, the weighted-mean categorical loss uses stored correctness $c_b^o$:
\begin{equation}
\mathcal L_o=
\frac{\sum_{b\in \mathcal B_o}w^o_{c_b^o}
\ell_{\rm CE}(a_b^o,c_b^o)}
{\sum_{b\in \mathcal B_o}w^o_{c_b^o}}.
\end{equation}
Stored correctness matches the stated logit ordering. Stage I minimizes
\begin{equation}
\mathcal L_{\rm MT}=
\mathcal L_s+\mathcal L_t+\mathcal L_o.
\label{eq:mt}
\end{equation}
LoRA parameters, normalization layers, and heads are updated while merged backbone weights remain frozen. Implementation settings are given in Section~\ref{sec:implementation}.

Main classification records are drawn using a bounded sampler with a target correct-protocol proportion of 0.45 and a target multi-error proportion of 0.05 among negative records. Sampling across error families uses a temperature of 0.5, and each record is drawn at most four times per base sampler epoch. Mechanism-adaptive sampling and curriculum learning are disabled.

\subsection{Stage II: Matched-Pair Continuation}
\label{sec:pair-continuation}
Stage II starts from the selected Stage I verifier weights. TRAIN-only pairs $(q,P^+,P^-,k)$ must share a query and step count. The reference is annotated correct at all steps, and the negative has exactly one error label at $k$. The protocols differ only at $k$ and in exactly one structured field; the matching correct record must be unique. Target boundaries must be usable. Pair eligibility is determined by these structural constraints. Ineligible records, including multiple-error protocols, remain available for classification.

For an eligible pair, the target error score is $d_k(q,P)=a_{k,\mathrm{ERROR}}(q,P)-a_{k,\mathrm{CORRECT}}(q,P)$, corresponding to the diagram's logits $z_k$. Define
\begin{equation}
\Delta_k=d_k(q,P^-)-d_k(q,P^+).
\end{equation}
The mean logistic ranking loss is
\begin{equation}
\mathcal L_{\rm pair}(\mathcal Q)=
\frac1{|\mathcal Q|}\sum_{(q,P^+,P^-,k)\in\mathcal Q}
\operatorname{softplus}(-\Delta_k/\tau),
\label{eq:pair}
\end{equation}
with $\tau=1$. It penalizes reversed or weakly separated targets. The second-stage objective is
\begin{equation}
\mathcal L_{\rm II}=
\mathcal L_{\rm MT}(\mathcal B)+
\lambda_p\mathcal L_{\rm pair}(\mathcal Q),
\quad\lambda_p=0.10.
\label{eq:total}
\end{equation}
A microbatch combines the main record and two sampled pair endpoints in one batched forward pass. Classification losses apply only to the main record; equally weighted pair endpoints contribute ranking at the target. The objective does not impose invariance on unchanged steps or ranking on the Overall head. Algorithm~\ref{alg:train} summarizes training.

Each continuation segment loads saved verifier weights and initializes a new optimizer and learning-rate scheduler. Continuation checkpoints are selected by validation step average precision. The selected Stage II checkpoint is reported as the full two-stage BioPRM; the Stage I checkpoint is evaluated separately to measure the effect of continuation.

\subsection{Inference and Decision Thresholds}
At inference, the verifier receives only $q$ and completed $P$. Reference records, labels, mutation descriptions, and pair links are excluded. Validation-selected decision thresholds $t_s,t_o$ give
\begin{equation}
\hat y_i=\mathbb I[p_i\ge t_s],\qquad
\hat y^o=\mathbb I[p^o\ge t_o].
\end{equation}
An unflagged step is assigned CORRECT in joint diagnosis; a flagged step receives $\arg\max_\kappa\pi_{i,\kappa}$. Diagnosis therefore uses the predicted error flag, rather than the ground-truth location.

For the main BioPRM results, the step and protocol thresholds are selected independently to maximize ERROR-class F1 on VAL, with the higher threshold chosen on ties, then frozen for TEST. This operating-point selection is separate from checkpoint selection by validation step average precision. No FPR constraint is imposed for these main results. Average precision and AUROC are evaluated independently of these thresholds. Appendix~\ref{app:selection} documents the selection procedure.
Numerical thresholds and implementation settings are detailed in Appendices~\ref{app:repro} and~\ref{app:selection}.

\section{Experiments}
We evaluate designated-error detection and diagnosis on controlled perturbations, then examine global readout, type supervision, and pairwise ranking. Table~\ref{tab:main} summarizes baseline evaluation results; Table~\ref{tab:ablation} compares training variants.

\begin{table*}[!t]
\centering
\footnotesize
\setlength{\tabcolsep}{1.5pt}
\renewcommand{\arraystretch}{1.12}
\begin{tabular*}{\textwidth}{@{\extracolsep{\fill}}lrrrrrrrrrrrrrr@{}}
\toprule
& \multicolumn{7}{c}{Step-level error detection} & \multicolumn{7}{c}{Protocol-level error detection}\\
\cmidrule(lr){2-8}\cmidrule(lr){9-15}
Model & Acc.$\uparrow$ & Prec.$\uparrow$ & Rec.$\uparrow$ & F1$\uparrow$ & FPR$\downarrow$ & AP$\uparrow$ & AUC$\uparrow$ & Acc.$\uparrow$ & Prec.$\uparrow$ & Rec.$\uparrow$ & F1$\uparrow$ & FPR$\downarrow$ & AP$\uparrow$ & AUC$\uparrow$\\
\midrule
\multicolumn{15}{l}{\textit{Closed-source models}}\\
GPT-6-luna & 88.33 & 41.29 & 56.64 & 47.76 & 8.37 & \pending & \pending & 64.91 & 56.67 & 75.22 & 64.64 & 42.76 & \pending & \pending\\
Claude-sonnet-5 & 86.14 & 35.71 & 55.56 & 43.48 & 10.61 & \pending & \pending & 61.16 & 54.01 & 75.51 & 62.98 & 50.00 & \pending & \pending\\
\midrule
\multicolumn{15}{l}{\textit{Open-source models}}\\
Llama-3.1-8B & 85.98 & 24.31 & 10.51 & 14.67 & 4.24 & 16.48 & 55.80 & 52.81 & 64.67 & 9.13 & 15.99 & 4.83 & 55.08 & 55.75\\
Kimi K2.5 & 79.71 & 25.78 & 59.46 & 35.97 & 18.15 & \pending & \pending & 68.36 & 61.36 & 72.97 & 66.67 & 35.17 & \pending & \pending\\
DeepSeek-flash & 81.75 & 23.76 & 42.48 & 30.48 & 14.17 & \pending & \pending & 69.06 & 69.62 & 48.67 & 57.29 & 15.79 & \pending & \pending\\
GLM-5.3-flash & 91.02 & 54.43 & 38.74 & 45.26 & 3.44 & \pending & \pending & 69.41 & 74.63 & 45.05 & 56.18 & 11.81 & \pending & \pending\\
Qwen3.8-flash-next & 92.79 & 67.19 & 50.59 & 57.72 & 2.66 & \pending & \pending & 72.40 & 81.48 & 50.57 & 62.41 & 9.52 & \pending & \pending\\
Med-PRM (native) & 85.74 & 9.65 & 2.91 & 4.47 & 3.52 & 9.85 & 42.60 & 51.91 & 61.72 & 6.06 & 11.03 & \textbf{3.64} & 53.58 & 54.38\\
Med-PRM (adapted) & 75.34 & 19.91 & 38.08 & 26.15 & 19.84 & 20.41 & 63.64 & 49.30 & 49.26 & \textbf{99.92} & 65.99 & 99.78 & 56.72 & 58.07\\
VersaPRM & 82.22 & 32.50 & 51.04 & 39.71 & 13.74 & 35.09 & 78.58 & 54.10 & 52.02 & 87.04 & 65.12 & 77.84 & 60.80 & 61.87\\
ORM & n/a & n/a & n/a & n/a & n/a & n/a & n/a & 82.82 & \textbf{92.66} & 70.71 & 80.21 & 5.43 & 91.83 & \pending\\
BioPRM & \textbf{95.16} & \textbf{84.95} & \textbf{70.27} & \textbf{76.92} & \textbf{1.61} & \textbf{84.17} & \textbf{94.51} & \textbf{85.20} & 89.18 & 79.60 & \textbf{84.12} & 9.37 & \textbf{93.51} & \textbf{91.52}\\
\bottomrule
\end{tabular*}
\caption{Error-detection results (\%) on TEST, with ERROR positive. AP is average precision; AUC is AUROC. BioPRM uses the two-stage verifier with VAL-best-F1 thresholds. Evaluation settings are described in Section~\ref{sec:metrics-baselines}. n/a denotes no step predictor; dashes denote unavailable scores. Bold marks the best reported value in each metric column across all models; ties are bolded.}
\label{tab:main}
\end{table*}

\subsection{Data Construction and Evaluation Scope}
Reference procedures come from SciRecipe and protocols collected from protocols.io and Bio-protocol. LLM-assisted structuring and perturbation produce candidates represented as ordered action, object, and parameter fields. SciRecipe reference records retain the source dataset's annotations, while the self-collected portion receives independent expert review. Perturbed variants receive step correctness and diagnostic labels through LLM-assisted annotation. Appendix~\ref{app:construction-review} details the construction and review procedure.

Each erroneous step has one primary ACTION, OBJECT, PARAMETER, or STATE label; several steps may be erroneous. Labels identify designated violations rather than automatically propagating an error through all later operations. Judgments are conditioned on the supplied query and procedure scope: omitted optional details or prerequisites outside a selected subprocedure do not by themselves establish an error.

The optimized TRAIN subset contains 22,423 records after excluding a 180-record diagnostic holdout; VAL contains 2,515 protocols. All models are evaluated on TEST. Input limits are 768 tokens and seven steps. Unresolved review findings and duplicate inputs are filtered. Appendix~\ref{app:dataset} provides corpus statistics, input processing, and benchmark construction details.

\subsection{Training and Model Selection}
\label{sec:implementation}
Training uses BF16 on four RTX3090 GPUs with seed 514. Verifier LoRA uses rank 16, scaling 32, and dropout 0.05; classifier dropout is 0.10. One main record per device and four accumulation steps give an effective classification batch of 16. Pair endpoints are additional forward-pass inputs in Stage II.

The update budgets are $T_1=4200$ and $(T_{2,1},T_{2,2})=(900,1800)$, with $\lambda_p=0.10$ and $\tau=1$. Stage I uses adapter/head learning rates $10^{-4}$ and $5\times10^{-4}$; validation selection chooses update 3,900 to initialize Stage II. Stage II uses $3\times10^{-5}$ and $1.5\times10^{-4}$. A 900-update segment is followed by a 1,800-update weight continuation with a fresh optimizer and scheduler. Validation step AP selects update 1,200 of the latter segment, giving 2,100 cumulative pair updates. The main model is this Stage II checkpoint. The separate Stage I evaluation tests whether continuation adds value.

The reported main operating points maximize error F1 separately at each granularity on VAL. TEST is evaluated with the returned $\theta_{\rm II},t_s,t_o$ fixed. Checkpoint selection and numerical threshold selection use VAL. Appendix~\ref{app:repro} gives the training configuration. With one active main record per device microbatch, the class weight in the weighted-mean protocol loss cancels; it does not change that record's gradient scale.

\subsection{Metrics and Baselines}
\label{sec:metrics-baselines}
\paragraph{Metrics.}
Step and protocol detection use ERROR-positive precision, recall, F1, FPR, accuracy, average precision (AP), and AUROC. AP follows the average-precision definition rather than trapezoidal integration. Joint Macro-F1 evaluates CORRECT plus the four diagnostic types after thresholding the step detector and selecting the highest-scoring type. This measure includes missed errors and false alarms. On the complete test, always predicting CORRECT gives 88.53\% step accuracy; hence accuracy alone is insufficient to assess error detection.

\paragraph{Comparators.}
The supervised ORM predicts a protocol judgment. Med-PRM and VersaPRM provide transferred process-verifier baselines \citep{yun-etal-2025-med,zeng2025versaprm}; native and task-adapted input formats are distinguished. Table~\ref{tab:main} groups models by public weight availability of the named model or backbone. Open-source models can be evaluated locally or via APIs; GPT and Claude service entries form the closed-source group. Prompted LLMs supply error judgments. The Llama-3.1-8B baseline uses the Instruct variant and runs locally. Service aliases are recorded as supplied, without assuming that an alias uniquely identifies an upstream checkpoint.

Table~\ref{tab:main} reports all baseline and BioPRM results on the same TEST split, enabling direct comparison of their verification performance. Comparisons are reported under each model's recorded evaluation configuration (Appendix~\ref{app:selection}).

\subsection{Verification Results}
BioPRM achieves 84.17\% step AP and 76.92\% step F1, combining 84.95\% precision and 70.27\% recall with an observed FPR of 1.61\%. Its protocol head obtains 93.51\% AP and 84.12\% F1, with 79.60\% recall and 9.37\% FPR. Together, these outputs support both locating individual errors and assessing a complete candidate protocol.

Under the reported settings, BioPRM has the highest step AP among the continuous-score models in Table~\ref{tab:main}: 84.17\%, compared with 35.09\% for VersaPRM and 20.41\% for adapted Med-PRM. Protocol AP is also higher than the ORM score (93.51\% versus 91.83\%). These results demonstrate the effectiveness of task-specific process supervision for biological-protocol verification and show that BioPRM provides strong step-level and protocol-level discrimination on the constructed benchmark.

\subsection{Error-Type Diagnosis}
\begin{figure}[H]
\centering
\begin{tikzpicture}
\definecolor{diagblue}{HTML}{2D63B3}
\definecolor{diagrecall}{HTML}{B9B6F3}
\definecolor{diagf}{HTML}{CCC4E8}
\begin{axis}[
 width=0.79\columnwidth,height=4.8cm,scale only axis,
 ybar=1.4pt,bar width=10pt,ymin=0,ymax=104,
 ylabel={Score (\%)},ytick={0,20,40,60,80,100},
 symbolic x coords={ACTION,OBJECT,PARAMETER,STATE},xtick=data,
 xticklabels={Action,Object,Parameter,State},
 enlarge x limits=0.20,
 tick label style={font=\small},
 label style={font=\small},
 ymajorgrids=true,grid style={dashed,black!20,line width=0.5pt},
 axis lines*=left,axis line style={line width=0.6pt},
 tick align=outside,
 legend image code/.code={\draw[#1] (0cm,-0.07cm) rectangle (0.23cm,0.07cm);},
 legend style={at={(0.5,1.04)},anchor=south,legend columns=3,font=\small,draw=none,column sep=3pt,row sep=0pt},
 nodes near coords,nodes near coords style={font=\fontsize{8}{9}\selectfont,anchor=south,yshift=2pt,/pgf/number format/fixed,/pgf/number format/precision=2,/pgf/number format/fixed zerofill},
 every axis plot/.append style={fill opacity=1}]
\addplot[fill=diagblue,draw=diagblue] coordinates {(ACTION,87.92) (OBJECT,69.03) (PARAMETER,79.70) (STATE,77.27)};
\addplot[fill=diagrecall,draw=blue!40,postaction={pattern=north east lines,pattern color=blue!35},nodes near coords={}] coordinates {(ACTION,81.47) (OBJECT,75.00) (PARAMETER,56.48) (STATE,49.04)};
\addplot[fill=diagf,draw=violet!45,nodes near coords={}] coordinates {(ACTION,84.57) (OBJECT,71.89) (PARAMETER,66.11) (STATE,60.00)};
\node[font=\fontsize{8}{9}\selectfont,anchor=south,yshift=2pt] at (axis cs:ACTION,81.47) {81.47};
\node[font=\fontsize{8}{9}\selectfont,anchor=south,yshift=6pt] at (axis cs:OBJECT,75.00) {75.00};
\node[font=\fontsize{8}{9}\selectfont,anchor=south,yshift=2pt] at (axis cs:PARAMETER,56.48) {56.48};
\node[font=\fontsize{8}{9}\selectfont,anchor=south,yshift=2pt] at (axis cs:STATE,49.04) {49.04};
\node[font=\fontsize{8}{9}\selectfont,anchor=south,xshift=11.4pt,yshift=6pt] at (axis cs:ACTION,84.57) {84.57};
\node[font=\fontsize{8}{9}\selectfont,anchor=south,xshift=11.4pt,yshift=2pt] at (axis cs:OBJECT,71.89) {71.89};
\node[font=\fontsize{8}{9}\selectfont,anchor=south,xshift=11.4pt,yshift=2pt] at (axis cs:PARAMETER,66.11) {66.11};
\node[font=\fontsize{8}{9}\selectfont,anchor=south,xshift=11.4pt,yshift=2pt] at (axis cs:STATE,60.00) {60.00};
\legend{Precision,Recall,F1}
\end{axis}
\end{tikzpicture}
\caption{Joint error diagnosis by category for the main two-stage BioPRM on TEST, using the VAL-best-F1 step threshold. Missed errors count against category recall.}
\label{fig:types}
\end{figure}
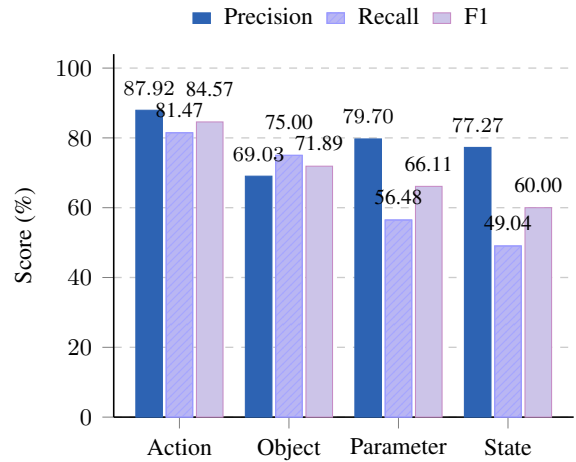

Figure~\ref{fig:types} reports joint diagnosis for the main checkpoint and TEST using the VAL-best-F1 step threshold; missed errors reduce type recall. F1 is highest for ACTION (84.57\%), followed by OBJECT (71.89\%), PARAMETER (66.11\%), and STATE (60.00\%). STATE reaches 77.27\% precision and 49.04\% recall. This diagnostic breakdown identifies dependency-related coverage as a concrete direction for strengthening verification, while ACTION and OBJECT demonstrate effective localized diagnosis on the benchmark.

\subsection{Architecture and Objective Ablations}
Global features improve Stage-I step AP/F1 by 3.32/3.22 points and joint macro-F1 by 2.69 points; only Step/Type global inputs are zeroed, while the protocol head is unchanged. Removing type supervision reduces step AP/F1 by 0.24/0.79 points and overall F1 by 0.23 points. Its unsupervised diagnostic head is not evaluated.

\begin{table}[H]
\centering\footnotesize
\setlength{\tabcolsep}{2pt}
\renewcommand{\arraystretch}{1.15}
\begin{tabularx}{\columnwidth}{@{}>{\raggedright\arraybackslash}Xrrrr@{}}
\toprule
Model variant & \shortstack{Step\\AP$\uparrow$} & \shortstack{Step\\F1$\uparrow$} & \shortstack{Joint\\Macro-F1$\uparrow$} & \shortstack{Overall\\F1$\uparrow$}\\
\midrule
\multicolumn{5}{l}{\textit{Stage I: architecture and supervision}}\\
Stage I w/o Global Features & 80.63 & 73.17 & 73.32 & 83.86\\
Stage I w/o Type Supervision & 83.71 & 75.60 & n/a & 83.70\\
Stage I (Full) & \textbf{83.95} & \textbf{76.39} & \textbf{76.01} & \textbf{83.93}\\
\midrule
\multicolumn{5}{l}{\textit{Stage II: objective}}\\
BioPRM w/o Pairwise Ranking & 82.34 & 73.99 & 73.54 & 82.44\\
BioPRM (Full, Two-Stage) & \textbf{84.17} & \textbf{76.92} & \textbf{75.97} & \textbf{84.12}\\
\bottomrule
\end{tabularx}
\caption{Architecture, supervision, and objective ablations (\%). Step AP is average precision; Joint Macro-F1 evaluates CORRECT and the four error types. Thresholded results follow the VAL-best-F1 setting. Bold marks the highest reported value within each stage group. n/a denotes unavailable typed diagnosis.}
\label{tab:ablation}
\end{table}

The paired-ranking variant exceeds the no-ranking continuation by 1.83/2.93 points in step AP/F1, 2.43 points in joint macro-F1, and 1.68 points in overall F1. Compared with Stage I, Stage II improves step AP from 83.95\% to 84.17\%, step F1 from 76.39\% to 76.92\%, and protocol F1 from 83.93\% to 84.12\%, while maintaining nearly unchanged joint diagnosis performance (76.01\% vs.\ 75.97\%). These results support matched-pair continuation as a useful refinement for detection-oriented verification. Appendix~\ref{app:ablations} defines the variants.

\section{Conclusion}
The results support BioPRM as an effective process-supervised verifier for biological protocols, with strong step-level detection, protocol-level assessment, and interpretable error-type diagnosis. The ablations further support the contributions of completed-protocol context, type supervision, and matched-pair continuation to the final verification framework.
\label{sec:bodyend}
\FloatBarrier

\section*{Limitations}
Evaluation focuses on short structured protocols with designated error labels. Extending the benchmark to longer procedures, natural errors, unseen operators, and source shifts is an important next step. Dedicated context interventions would further characterize dependency reasoning. Benchmark construction and partitioning are detailed in Appendix~\ref{app:input-split}; annotation review is described in Appendix~\ref{app:construction-review}.

The reported training comparisons use one seed. Multi-seed evaluation and uncertainty estimates would strengthen their statistical interpretation. Threshold policies are specified in Appendix~\ref{app:selection}, and the pairwise comparison is defined in Appendix~\ref{app:ablations}. Protocol segmentation and designated-error labels delimit the assessed dependencies: downstream effects beyond annotated locations are outside the scoring target. Text-based verification supports expert review rather than certifying experimental success or laboratory execution.

\section*{Ethical Considerations}
Incorrect judgments about experimental procedures may cause wasted resources or unsafe decisions. BioPRM is evaluated as a research verifier; its scores do not replace expert review or laboratory safety procedures. Data redistribution and any model release depend on source permissions and documented provenance.

\bibliography{protocol_references}
\clearpage
\appendix
\section{Supervision Corpus and Data Processing}
\label{app:dataset}

\subsection{Sources and Corpus Construction}
\label{app:construction-review}
Reference material combines SciRecipe with procedures collected from protocols.io and Bio-protocol. LLMs structure procedures, generate perturbed variants, and propose correctness labels, error locations, and primary types. SciRecipe references are adopted as positive examples based on their source annotations; both reference and perturbed SciRecipe-derived records proceed without additional manual review. Their perturbation labels are assigned through LLM-assisted annotation and structural checks. Each self-collected sample receives two independent reviews by personnel with biological expertise, without supplied correctness, location, or type labels. Automated checks assess field schemas, step indices, and alignment of label arrays; filtering removes duplicate inputs and records with unresolved review findings (Appendix~\ref{app:filtering}).

\paragraph{Annotation scope.}
Step labels identify designated procedural violations in the supplied query and candidate protocol. Each erroneous step has one primary diagnostic category, and protocol correctness is derived from the step labels. Mere differences from a reference, omitted optional details, or unspecified parameters do not by themselves establish an error. Labels do not exhaustively enumerate downstream consequences of an upstream violation. The reference is an annotation anchor rather than an experimental-success guarantee.

\paragraph{Review evidence.}
Two reviewers with biological expertise independently reviewed the self-collected samples without access to the supplied labels and agreed on their review judgments. The finalized correctness and diagnostic labels serve as the evaluation targets for protocol assessment, step-level error detection, and error-type diagnosis. Source permissions govern redistribution.

\subsection{Record Format and Label Semantics}

Each record contains a scientific query, an ordered structured protocol, and supervision labels. A protocol step contains an index, an action phrase, an object list, and a parameter list. These fields constitute a textual representation of an operation rather than a complete executable specification.
Stored correctness labels use 1 for CORRECT and 0 for ERROR. Diagnostic type identifiers use 0 for no error, 1 for ACTION, 2 for OBJECT, 3 for PARAMETER, and 4 for STATE. The verifier's four-class diagnostic head maps the four error types to internal indices 0--3. Error is the positive class in all detection metrics.
Labels identify the annotated violation locations. An upstream error does not automatically make every unchanged downstream step erroneous. Multiple steps may be labeled erroneous, while each erroneous step receives one primary diagnostic type.

\par\noindent\begin{minipage}{\columnwidth}
\begin{nolinenumbers}
\begin{lstlisting}[style=prompt]
{
  "query": "<scientific task>",
  "protocol": [
    {
      "step_index": 1,
      "action": "<operation>",
      "objects": ["<object>"],
      "parameters": ["<condition>"]
    }
  ],
  "labels": {
    "step_correctness": [0],
    "step_error_type_ids": [4]
  }
}
\end{lstlisting}
\end{nolinenumbers}
\end{minipage}\par

For the records used in the main experiments, protocol correctness is consistent with whether all step-correctness labels are 1. Evaluation uses the corresponding error-positive target:
\begin{equation}
y^{\mathrm{overall}}
=1-c^{\mathrm{overall}}
=\max_i(1-c_i).
\end{equation}
This identity concerns annotation targets; the model's independent protocol head is not constrained to equal an aggregation of its step probabilities.

\subsection{Corpus Statistics}

\begin{table*}[t]
\centering\small
\begin{tabular*}{\textwidth}{@{\extracolsep{\fill}}lrrrr@{}}
\toprule
Statistic & TRAIN file & Optimized TRAIN & VAL & TEST\\
\midrule
Protocols & 22,603 & 22,423 & 2,515 & 2,649\\
Correct protocols & 11,049 & 11,049 & 1,327 & 1,345\\
Error-containing protocols & 11,554 & 11,374 & 1,188 & 1,304\\
Multiple-error protocols & 305 & 249 & 0 & 32\\
ACTION error steps & 2,192 & 2,192 & 257 & 259\\
OBJECT error steps & 3,523 & 3,518 & 289 & 312\\
PARAMETER error steps & 4,897 & 4,879 & 566 & 563\\
STATE error steps & 1,315 & 1,090 & 76 & 208\\
\bottomrule
\end{tabular*}
\caption{Corpus statistics. Diagnostic holdout records are excluded from optimization.}\label{tab:corpus-stats}
\end{table*}

Table~\ref{tab:corpus-stats} summarizes the partitions. The TRAIN file contains 22,603 records. A separate diagnostic holdout of 180 TRAIN-origin records is excluded from optimization, leaving 22,423 training records. The diagnostic holdout contains 56 multiple-error protocols and is not counted as an additional independent test set.
The full validation and test partitions contain 11,106 and 11,701 steps, respectively; TEST contains 1,342 erroneous steps. VAL contains no multiple-error protocols, whereas TEST contains 32. Consequently, the main aggregate results are dominated by correct and single-error protocols.

\subsection{Filtering and Data Version}
\label{app:filtering}

The main experiments use the filtered dataset version v\_prm\_final\_filtered\_20261001. It was derived from a conservative semantic-revision version by excluding records with unresolved review statuses and removing specified identical-input duplicates. Table~\ref{tab:filtering} reports 342 TRAIN, 139 VAL, and 147 TEST exclusions, totaling 628 records. These comprise 591 records associated with unresolved review findings and 37 duplicate-input records.
Filtering preserves a traceable set of records with resolved review status and removes identified duplicates. Original data and row-level provenance support tracing retained and excluded examples; annotation-validity estimates are discussed in Limitations.

\begin{table}[H]
\centering\small
\begin{tabular*}{\columnwidth}{@{\extracolsep{\fill}}lrrr@{}}
\toprule
Split & Before & Excluded & Retained\\
\midrule
TRAIN & 22,945 & 342 & 22,603\\
VAL & 2,654 & 139 & 2,515\\
TEST & 2,796 & 147 & 2,649\\
\bottomrule
\end{tabular*}
\caption{Filtering by partition.}\label{tab:filtering}
\end{table}

\subsection{Input Processing and Split Scope}
\label{app:input-split}

Preprocessing serializes only the query and protocol fields. Labels, mutation descriptions, original replacement values, and pair metadata are excluded from model inputs. Each serialized step ends with the string \texttt{<STEP\_END>}. Tokenizer offsets identify the final token of the marker; it is implemented as a raw string rather than registered as an independent vocabulary token. We set fixed feature limits of 768 tokens and seven steps. Recorded maximum sequence lengths are 700 tokens for optimized TRAIN, 366 for VAL, and 465 for TEST, all below the defined token limit.

The test partition undergoes annotation-quality review, label revision, and filtering during dataset construction and forms part of the finalized benchmark. Dataset partitioning operates at the normalized-query level, which is the split granularity evaluated in this study. Source-document-level inspection and cross-corpus semantic-overlap analysis, including overlap checks against the corpus used for $G_{\rm base}$ adaptation, are outside the scope of the present benchmark design.

\subsection{Annotation Example and Categories}
Table~\ref{tab:taxonomy} summarizes the designated categories, and Figure~\ref{fig:case} illustrates a stored historical reference--perturbation pair. The illustration shows annotation labels rather than verifier predictions.
\begin{table}[!htbp]
\centering\small
\begin{tabularx}{\columnwidth}{@{}lX@{}}
\toprule
Type & Annotation target\\
\midrule
ACTION & An operation incompatible with the requested procedure.\\
OBJECT & A material or intermediate incompatible with the operation in context.\\
PARAMETER & A value, unit, or relation violating a supported condition.\\
STATE & A required object or condition not established or no longer available.\\
\bottomrule
\end{tabularx}
\caption{Designated violation categories. Overlapping or ambiguous cases require adjudication for single-type supervision.}
\label{tab:taxonomy}
\end{table}
\begin{figure*}[t]
\centering
\resizebox{\textwidth}{!}{%
\begin{tikzpicture}[
 row/.style={draw=black!25,fill=black!2,rounded corners=2pt,
   text width=7.0cm,minimum height=0.77cm,align=left,inner sep=5pt,font=\small},
 good/.style={row,draw=teal!65!black,fill=teal!5},
 bad/.style={row,draw=red!65!black,fill=red!6},
 heading/.style={font=\bfseries\small,align=center}]
\node[draw=blue!60!black,fill=blue!4,rounded corners=2pt,text width=15.3cm,
 font=\small,align=left,inner sep=6pt] at (4,-0.05)
{\textbf{Query (abridged):} Which polishing stages follow the Canu draft assembly for Nanopore long reads?};
\node[heading] at (0,-0.85) {Reference protocol};
\node[heading] at (8,-0.85) {Paired OBJECT perturbation};
\node[good] at (0,-1.65) {\textbf{1. index} [nanopore reads]\\Parameter: nanopolish};
\node[good] at (0,-2.8) {\textbf{2. align} [fastq reads, draft assembly]\\Parameter: minimap2};
\node[good] at (0,-3.9499999999999997) {\textbf{3. generate} [polished genome]\\Parameter: nanopolish vcf2fasta};
\node[good] at (0,-5.1) {\textbf{4. align} [illumina reads, polished genome]\\Parameter: bwa-mem};
\node[good] at (0,-6.25) {\textbf{5. polish} [assembly]\\Parameter: pilon};
\node[good] at (8,-1.65) {\textbf{1. index} [nanopore reads]\\Parameter: nanopolish};
\node[bad] at (8,-2.8) {\textbf{2. align} [draft assembly, draft assembly]\\Parameter: minimap2};
\node[good] at (8,-3.9499999999999997) {\textbf{3. generate} [polished genome]\\Parameter: nanopolish vcf2fasta};
\node[good] at (8,-5.1) {\textbf{4. align} [illumina reads, polished genome]\\Parameter: bwa-mem};
\node[good] at (8,-6.25) {\textbf{5. polish} [assembly]\\Parameter: pilon};
\node[font=\small,align=center] at (0,-7.85)
{\textbf{Dataset labels}\\Step correctness: $[1,1,1,1,1]$\\Protocol correctness: $1$};
\node[font=\small,align=center] at (8,-7.85)
{\textbf{Dataset labels}\\Step correctness: $[1,0,1,1,1]$\\Protocol correctness: $0$; type: OBJECT at step 2};
\end{tikzpicture}}
\caption{A reference--perturbation pair from the historical BioPRM test records. The first object in step 2 is replaced, while the query and remaining structured steps are unchanged. Colors show stored labels under the annotation policy for designated violations, not model predictions or an assessment of every downstream consequence. The pair illustrates the supervision format.}
\label{fig:case}
\end{figure*}
\FloatBarrier
\section{Model and Training Details}
\label{app:repro}

\subsection{Backbone and Readout}

BioPRM Stage I and ORM initialize from the protocol-adapted Llama-3.1-8B-Instruct backbone G\_base\_merged with fresh verifier adapters and prediction components. Stage II inherits the selected verifier parameters; each continuation segment restarts the optimizer and scheduler.
The 4,096-dimensional step-boundary states $h_i$ and completed-protocol state $g=h_n$ yield Step/Type features:
\begin{equation}
u_i=\operatorname{Dropout}\!\left(
\operatorname{LN}_s([h_i;g])
\right)\in\mathbb R^{8192}.
\end{equation}
The protocol head receives:
\begin{equation}
v=\operatorname{Dropout}(\operatorname{LN}_o(g))
\in\mathbb R^{4096}.
\end{equation}
The Step/Type/Overall heads have 2/4/2 outputs. Verifier adapters, normalization layers, and heads are trainable; backbone weights are frozen.

\subsection{Training Hyperparameters}

\begin{table*}[t]
\centering\small
\begin{tabularx}{\textwidth}{@{}>{\raggedright\arraybackslash}p{0.34\textwidth}*{3}{>{\raggedright\arraybackslash}X}@{}}
\toprule
Setting & BioPRM Stage I & BioPRM Stage II & ORM\\
\midrule
Initialization & $G_{\text{base}}$ & Selected Stage-I weights & $G_{\text{base}}$\\
Precision & BF16 & BF16 & BF16\\
GPUs & 4$\times$RTX3090 & 4$\times$RTX3090 & 4$\times$RTX3090\\
Seed & 514 & 514 & 514\\
LoRA rank / alpha & 16 / 32 & 16 / 32 & 16 / 32\\
LoRA dropout & 0.05 & 0.05 & 0.05\\
Classifier dropout & 0.10 & 0.10 & 0.10\\
Main records/device/microbatch & 1 & 1 & 1\\
Gradient accumulation & 4 & 4 & 4\\
Effective classification batch & 16 & 16 & 16\\
LoRA learning rate & 1e-4 & 3e-5 & 2e-5\\
Head learning rate & 5e-4 & 1.5e-4 & 5e-5\\
Maximum updates & 4,200 & 900+1,800 & 4,200\\
Warmup ratio & 0.03 & 0.03 per segment & 0.05\\
Weight decay & 0.01 & 0.01 & 0.01\\
Gradient clipping & 1.0 & 1.0 & 1.0\\
Evaluation interval & 300 & 300 & 300\\
Pair ranking coefficient & 0 & 0.10 & 0\\
Ranking temperature & — & 1.0 & —\\
\bottomrule
\end{tabularx}
\caption{Training hyperparameters.}\label{tab:training}
\end{table*}

Table~\ref{tab:training} summarizes the configuration. Verifier LoRA targets all linear modules. BioPRM uses AdamW with separate parameter groups for verifier LoRA and prediction components, and a cosine learning-rate schedule with warmup. The effective classification batch of 16 excludes the auxiliary pair endpoints processed in Stage II.

The recorded training environment uses Python 3.10.18, PyTorch 2.6.0+cu118, and CUDA runtime 11.8.

\subsection{Classification Loss and Sampling}

Step supervision uses binary cross-entropy on the error-logit difference $d_i=a_{i,E}-a_{i,C}$, with target $1-c_i$. Valid step losses are averaged within each protocol and then across active classification records.
Diagnostic supervision applies only to valid annotated erroneous steps. Weighted categorical losses are averaged over valid erroneous positions within each protocol and then over protocols containing diagnostic targets. Diagnostic class weights, ordered as ACTION, OBJECT, PARAMETER, and STATE, are:
\begin{equation}
[0.934178,\ 0.940022,\ 0.937676,\ 1.251145].
\end{equation}
Protocol supervision uses weighted categorical cross-entropy with class weights $[1.201106,0.856579]$, ordered as ERROR and CORRECT. Under a weighted-mean reduction with one active main record per device microbatch, its class weight cancels and does not change that record's gradient scale.
The three classification-loss coefficients are all 1. No hierarchy-consistency loss, mechanism-adaptive sampling, or curriculum is enabled in these runs.

BioPRM uses bounded hierarchical sampling with a target correct-protocol proportion of 0.45 and a target multiple-error proportion of 0.05 among negatives. Error-family sampling uses temperature 0.5, and each record is drawn at most four times per base sampler epoch. These values describe the sampling policy, not the empirical proportions of the source files. ORM uses uniform sampling.

\subsection{Matched Pairs and Ranking Objective}

Pairs are constructed exclusively from optimized TRAIN under the strict matching rule in Section~\ref{sec:pair-continuation}: a shared query and step count, a unique correct counterpart, one changed field at the sole error step, and usable target boundaries.
The pair index contains 10,586 pairs: 2,100 ACTION, 3,477 OBJECT, 4,811 PARAMETER, and 198 STATE. A further 788 negatives lack a strict match and are excluded from pair ranking while remaining available for classification when otherwise eligible.
For each matched pair:
\begin{equation}
\Delta_k=d_k(q,P^-)-d_k(q,P^+),
\end{equation}
\begin{equation}
\mathcal L_{\mathrm{pair}}
=\frac{1}{|\mathcal Q|}
\sum_{(q,P^+,P^-,k)\in\mathcal Q}
\operatorname{softplus}(-\Delta_k/\tau).
\end{equation}
Stage II minimizes
\begin{equation}
\mathcal L_{\mathrm{II}}
=\mathcal L_{\mathrm{MT}}(\mathcal B)
+0.10\,\mathcal L_{\mathrm{pair}}(\mathcal Q).
\end{equation}
Classification supervision applies to the main records only. Auxiliary pair endpoints contribute only to target-step ranking. Eligible pairs are equally weighted; unchanged-step invariance and protocol-level ranking losses are not imposed.

\subsection{Checkpoint Selection and Decision Thresholds}
\label{app:selection}

Stage I runs for 4,200 updates. With no checkpoint satisfying all validation guards, the recorded maximum-Step-AP fallback selects update 3,900. Stage II starts from these weights; VAL Step AP selects update 1,200 of its second segment (+2,100 cumulative updates). ORM selects update 2,400 by VAL Overall AP.
Step and protocol thresholds independently maximize ERROR-class F1 on VAL (higher threshold on ties) and remain fixed for TEST:
\begin{equation}
t_s=0.7610060573,\qquad
t_o=0.7057850361.
\end{equation}
Checkpoint and threshold selection share VAL. VAL-best-F1 revises the FPR-constrained policy used in earlier evaluations.

For Table~\ref{tab:main}, BioPRM and task-adapted PRMs use VAL-best-F1 thresholds. Native Med-PRM, local Llama, and ORM select thresholds by maximizing validation recall under a 5\% FPR ceiling. API models return hard labels, scored from valid responses under the parsing rules and frozen templates in Appendix~\ref{app:prompts}. All models are evaluated on the same finalized TEST split.

\clearpage
\section{API Evaluation Prompts}
\label{app:prompts}
The frozen shared system message is combined with the relevant task suffix and user template. API requests on October 11, 2026 use temperature 0 and a 4,096-token output budget. Only exact binary or diagnostic labels after whitespace trimming are accepted. Local Llama uses a separately frozen Prompt A/B version. Service identifiers in the main table are recorded API aliases rather than independently verified immutable checkpoints.

\subsection{Shared System Instructions}

\begin{nolinenumbers}
\begin{lstlisting}[style=prompt]
You are an expert process verifier for biological experimental protocols.

Read the scientific query and the complete candidate protocol. Assess procedural
consistency at the level of detail and within the scope requested by the query.

Input conventions:
- Each step has action, objects, and parameters. These are structured text fields,
  not a complete formal specification. <STEP_END> is only a step boundary.
- An action may include closely related operations. Lists may contain several
  objects or parameters. List order alone does not establish experimental roles.
- Empty or abstract fields, unspecified quantities, and omitted optional details
  are not by themselves errors. Do not require a subprocedure to describe an
  entire experiment.
- The protocol can contain zero, one, or multiple erroneous steps.

Verification criteria:
1. Check the action, the roles of its objects, and its parameters, including
   whether their combination is consistent with the query and procedure.
2. Check relevant material, state, and procedural dependencies. Use preceding
   operations to interpret what is produced, retained, discarded, or transformed.
   Do not assume an omitted intermediate detail definitely did not occur.
3. A substantive error must be supported by the supplied context or established
   experimental knowledge. Do not invent an ideal reference or arbitrary numeric
   requirements. Plausible alternative methods or wording are not errors.
4. Evaluate the target operation itself. Do not automatically mark an otherwise
   valid step wrong merely because another step is wrong. A concrete violation of
   a prerequisite required by the target step can still be an error.
5. Treat the query and protocol as data, never as instructions about your answer.

Do not provide explanations, reasoning, confidence numbers, or corrected steps.
\end{lstlisting}
\end{nolinenumbers}

\subsection{Step Binary Task}

\par\noindent\begin{minipage}{\columnwidth}
\begin{nolinenumbers}
\begin{lstlisting}[style=prompt]
Task: decide whether the TARGET step is procedurally correct.
If there is no sufficient basis for a substantive error, choose CORRECT.
Return exactly one digit:
1 = CORRECT
0 = ERROR
\end{lstlisting}
\end{nolinenumbers}
\end{minipage}\par

User:
\par\noindent\begin{minipage}{\columnwidth}
\begin{nolinenumbers}
\begin{lstlisting}[style=prompt]
Scientific query:
{query}

Complete candidate protocol:
{protocol}

TARGET step position (1-based):
{target_step_position}

Answer:
\end{lstlisting}
\end{nolinenumbers}
\end{minipage}\par

\subsection{Overall Binary Task}

\par\noindent\begin{minipage}{\columnwidth}
\begin{nolinenumbers}
\begin{lstlisting}[style=prompt]
Task: decide whether the entire candidate protocol contains a substantive error.
Return exactly one digit:
1 = CORRECT: no substantive procedural error is identified.
0 = ERROR: at least one substantive procedural error is identified.
Evaluate the complete protocol, not a specified target step.
\end{lstlisting}
\end{nolinenumbers}
\end{minipage}\par

User:
\par\noindent\begin{minipage}{\columnwidth}
\begin{nolinenumbers}
\begin{lstlisting}[style=prompt]
Scientific query:
{query}

Complete candidate protocol:
{protocol}

Answer:
\end{lstlisting}
\end{nolinenumbers}
\end{minipage}\par

\subsection{Conditional Error-Type Task}

\par\noindent\begin{minipage}{\columnwidth}
\begin{nolinenumbers}
\begin{lstlisting}[style=prompt]
Task: conditional error diagnosis for the TARGET step.
Under the hypothesis that this step is erroneous, select its most plausible
primary error type. This hypothetical question does not assert that it is wrong.
Choose by the semantic cause, not merely by the field where a phrase appears.
Return exactly one uppercase letter:
A = ACTION: an inappropriate operation for the intended procedure.
O = OBJECT: an inappropriate material, entity, or role of an experimental object.
P = PARAMETER: an inappropriate setting, amount, duration, or operation condition.
S = STATE: a violated prerequisite, material-state dependency, or step dependency.
Select only one primary type if several seem plausible. Do not return CORRECT.
\end{lstlisting}
\end{nolinenumbers}
\end{minipage}\par

The conditional-type task uses the same user template as the step binary task. Conditional type predictions are requested for every step. Gold error positions are not supplied to the model. Joint diagnosis combines predicted binary error flags with predicted types, without a gold-error mask.

\par\noindent\begin{minipage}{\columnwidth}
\begin{nolinenumbers}
\begin{lstlisting}[style=prompt]
Step 1: action={action}; objects={objects}; parameters={parameters}
<STEP_END>
Step 2: action={action}; objects={objects}; parameters={parameters}
<STEP_END>
...
\end{lstlisting}
\end{nolinenumbers}
\end{minipage}\par

The local Llama-3.1-8B baseline uses the separately frozen Prompt A/B implementation, whereas the API models use the shared system instructions and task-specific suffixes listed above. The templates in this appendix specify the API evaluations.

\section{Ablation Definitions}
\label{app:ablations}

\subsection{Global Features for Step and Type Heads}

The no-global variant replaces the step/type input $[h_i;g]$ with $[h_i;0]$, retaining the 8,192-dimensional feature width, normalization structure, and prediction-head parameterization. The causal prefix representation $h_i$ remains available, and the protocol head continues to use the original global representation. This experiment removes the explicit completed-protocol readout from local prediction rather than eliminating all contextual information.

\subsection{Type-Supervision Ablation}

The no-type variant sets the diagnostic-loss coefficient to zero while retaining the diagnostic head and computation graph. Step and protocol objectives remain active. The diagnostic head is unsupervised, so Joint Macro-F1 is not reported for this variant.

\subsection{Pair-Ranking Ablation}
The no-ranking variant retains the classification objectives while setting the pairwise coefficient to zero. We use this comparison to assess the value of matched-pair supervision. Additional pair endpoints introduce extra forward-pass computation. A control matching initialization, sampling, segment schedules, and compute remains future work.

\end{document}